\documentclass[twocolumn,floatfix,prl]{revtex4-2}%
\usepackage[dvipdfmx]{graphicx}%
\usepackage{amsmath}%
\usepackage{amsfonts}%
\usepackage{amssymb}
\usepackage{bm}
\usepackage{color}
\usepackage{ulem}

\def\e{{\epsilon}}
\def\k{{ {\bm k} }}
\def\p{{ {\bm p} }}
\def\q{{ {\bm q} }}
\def\Q{{ {\bm Q} }}

\def\0{{ {\bm 0} }}
\def\w{{\omega}}
\def\a{{\alpha}}

\allowdisplaybreaks[4]

\begin{document}
\title{
%Preempting 
Hidden antiferro-nematic order in 
Fe-based superconductor BaFe$_2$As$_2$ and NaFeAs
above $T_S$
}
\author{
Seiichiro Onari and Hiroshi Kontani
}

\date{\today }

\begin{abstract}
In several Fe-based superconductors, slight $C_4$ symmetry breaking occurs at $T^*$,
which is tens of Kelvin higher than the structural transition temperature $T_S$.
In this ``hidden'' nematic state at $T_S<T<T^*$, 
the orthorhombicity is tiny [$\phi=(a-b)/(a+b) \ll 0.1$\%],
but clear evidences of bulk phase transition have been accumulated.
To explain this long-standing mystery,
we propose the emergence of antiferro-bond (AFB) order
 with the antiferro wavevector $\q=(0,\pi)$ at $T=T^*$, by which the characteristic phenomena below $T^*$ are satisfactorily explained.
%Below $T^*$, the AFB order reproduces the pseudogap
%and the small nematicity $\psi\propto T^*-T$.
%The AFB order at $T^*$ does not interrupt the ferro-orbital order at
% $T_S$ thanks to the prominent orbital selectivity of nematicity.
This AFB order originates from the 
inter-orbital nesting between the $d_{xy}$-orbital 
hole-pocket and the electron-pocket, and this inter-orbital bond order
%with $d_{xy}$-orbital hole-pocket,
%due to the frustration between three $t_{2g}$ d-electrons.
naturally explains the pseudogap, band-folding, and tiny nematicity that
 is linear in $T^*-T$. 
%The AFB order at $T^*$ does not interrupt the ferro-orbital order at $T_S$ owing to the difference in the orbital selectivity.
%\Erase{In addition, we discuss the significant role of AFB fluctuations on the pairing mechanism in Ba122 families.}
 The hidden AFB order explains key experiments in both BaFe$_2$As$_2$ and NaFeAs, 
but it is not expected to occur in FeSe because of the absence of the $d_{xy}$-orbital hole-pocket.

\end{abstract}

\address{
 Department of Physics, Nagoya University,
Furo-cho, Nagoya 464-8602, Japan. 
}
 
%\pacs{74.70.Xa, 75.25.Dk, 74.20.Pq} 
%74.20.Pq Electronic structure calculations

\sloppy

\maketitle

%%%%%%%%%%%%%%%%%%
%Introduction
%%%%%%%%%%%%%%%%%%
%nematic electronic state

The emergence of rich nematic phase transitions is a central unsolved issue in Fe-based superconductors.
At the structural transition temperature $T_S$, ferro-orbital (FO) order with
$\psi\equiv (n_{xz}-n_{yz})/(n_{xz}+n_{yz})\ne 0$ is driven by electron
correlation \cite{ARPES}, by which the orthorhombicity
$\phi=(a-b)/(a+b)$ occurs in proportion to $\psi$.
Above $T_S$,
the electronic nematic susceptibility
develops divergently \cite{Yoshizawa,Bohmer,Gallais,Raman2}.
%The intimate relationship between nematicity and magnetism 
%has been discussed based on the 
As possible mechanisms of nematicity, both
spin-nematic scenarios
\cite{Fernandes,Fernandes-122,DHLee,QSi,Valenti,Fang,Fernandes-review}
and the orbital/charge-order scenarios
\cite{Kruger,PP,WKu,Kontani-Saito-Onari,Onari-SCVC,Onari-SCVCS,Onari-form,FeSe-Yamakawa,Text-SCVC,JP,Fanfarillo,Chubukov-RG}
have been proposed. Both scenarios were successfully applied to the
nematicity in BaTi$_2$Sb$_2$O
\cite{BaTi2Sb2O,Fernandes-BaTi2Sb2O,Nakaoka} and cuprate superconductors \cite{Tsuchiizu-Cu}.

\begin{figure}[!htb]
\includegraphics[width=.8\linewidth]{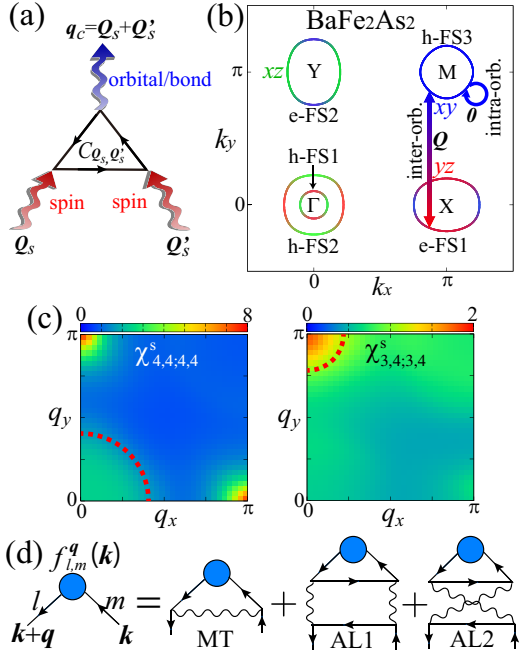}
\caption{
%(color online)
(a) Quantum process of the spin-fluctuation-driven orbital fluctuations with
 $\bm{q}_c=\Q_s+\Q'_s$.
(b) FSs of the BaFe$_2$As$_2$ model in the unfolded zone.
The colors green, red and blue correspond to 
orbitals 2, 3, and 4, respectively. 
%We depict the inter-orbital interaction due to the
% nesting $\Q=(0,\pi)$ between h-FS3 (orbital 4) and e-FS1 (orbital 3) as
% well as the forward intra-orbital interaction for
% orbital 4 at M.
(c) $\q$ dependences of 
 $\chi^{s}_{4,4;4,4}(\q,0)$ and $\chi^{s}_{3,4;3,4}(\q,0)$ given by the RPA. The large antiferro- and ferro-fluctuations shown by dotted circles are significant for the AFB order formation.
(d) Feynman diagrams of the DW equation. Each wavy line represents a
 fluctuation-mediated interaction.
%(g) Feynman diagrams of the four-point vertex $\hat{I}^{\bm{q}}(k,k')$ and $\hat{P}^{\bm{q}}(k,k')$.
}
\label{fig:FS}
\end{figure}

However, the nematicity in Fe-based superconductors recently
exhibits
very rich variety beyond the original expectation.
%very rich phase diagrams with nematicity and magnetism.
%The rich variety of phase diagram with nematicity and magnetism
%in Fe-based superconductors 
%has been attracting increasing attention.
%offers a significant hint to
%reveal the fundamental many-body electronic states.
Well-known discoveries are the nematicity without magnetization in FeSe and
the nematicity with $B_{2g}$ symmetry in the heavily hole-doped compound 
AFe$_2$As$_2$ (A=Cs, Rb)
\cite{CsFe2As2-nematic,RbFe2As2-nematic,Shibauchi-B2g,Onari-B2g,Fernandes-B2g}, which
is rotated by $45^\circ$ with respect to the nematicity in FeSe.
These nematic orders are naturally understood as the
ferro orbital and/or bond orders driven by the interference between spin
fluctuations described in Fig. \ref{fig:FS}(a) \cite{Onari-SCVC}, where
$C_{\Q_s,\Q'_s}$ gives the three-boson coupling. 
%spin-fluctuation-driven orbital-order or bond-order described by 
%the Aslamazov--Larkin (AL) in Fig. \ref{fig:FS}(a) and Maki--Thompson (MT) vertex corrections (VCs)
%\cite{Onari-SCVC}.

The most significant open issue in the nematicity 
is the emergence of another type of nematicity in various
Ba122 compounds below $T=T^*$, which is higher than $T_S$
by tens of Kelvin.
A true second-order bulk nematic transition at $T^*$ has been reported
in many experimental studies,
such as a magnetic torque study \cite{Kasahara-torque}, an X-ray study \cite{X-ray}, an optical measurement study \cite{Thewalt}, and
a laser
photoemission electron microscope study \cite{Shimojima-PEEM}.

Below $T^*$, the orthorhombicity $\phi$ is finite but very small ($\ll
0.1$\%), but a sizable pseudogap and shadow band exist \cite{Shimojima-PG,Shimojima-SB}.
The exponent of the nematicity $\psi\propto\phi\propto (T^*-T)^\a$
is $\a\sim 1$, which is much larger than the mean-field 
exponent ($1/2$). The relation $\phi\propto(T^*-T)$ is also observed in
NaFeAs \cite{NaFeAs-T*}.
One may consider that the nematicity at $T^*$
is not a true phase transition
but that it reflects the inhomogeneity of the FO-order transition 
temperature $T_S$ due to local uniaxial pressure and randomness
\cite{Fernandes-122,Dai,Dai2}.
On the other hand,
$T^*$ seems not to be sensitive to the sample quality, 
and the domain structure of nematicity observed in the $C_4$ phase above
$T_S$ \cite{Thewalt,Shimojima-PEEM} is homogeneous.
The aim of this study is to reveal the origin of this mysterious
hidden nematic state below $T=T^*$ and to explain why multistage-nematic transitions 
(at $T=T^*$ and $T_S$) emerge in Ba122 and NaFeAs families.

In this paper,
we predict the emergence of antiferro-bond (AFB) order with the antiferro
wavevector $\q=(0,\pi)$ at $T=T^*$ above the FO-order transition
temperature $T_S$.
%above the structure transition temperature $T_S$.
Below $T^*$, the AFB order causes a pseudogap in the density of states
and the small $T$-linear nematicity $\psi\propto T^*-T$.
The AFB order does not interrupt the ferro-orbital order at $T_S$, because these order parameters have different orbital components.
Thus, both the spin and nematic susceptibilities, 
$\chi^s(\Q)$ and $\chi_{\rm nem}({\bm 0})$ respectively,
 show only a small anomaly at $T=T^*$.
The obtained inter-orbital AFB order is driven by the interference
between antiferro and ferro spin fluctuations, which are caused by the
inter-orbital nesting between the $d_{xy}$-orbital hole-pocket and electron
pockets.
%The present AFB ordered state above $T_S$ may be
%realized in NaFeAs, where the $T$-linear behavior of nematicity is reported\cite{NaFeAs-T*}.
%However, the AFB order is absent in FeSe since
%the $d_{xy}$-orbital hole-pocket is absent.
The present theory naturally explains the 
long-standing mystery of the hidden nematic state below $T^*$ in Ba122
 and NaFeAs families, in both of which $T$-linear nematicity has been
 reported \cite{Kasahara-torque,NaFeAs-T*}. 

In contrast,
the AFB order does not occur in the FeSe model that has no
$d_{xy}$-orbital hole-pocket, since the inter-orbital nesting is essential to realize the
AFB order. This result is consistent with the absence of $T^*$ in FeSe \cite{Shimojima-FeSe}.
%\Add{We also predict that the AFB fluctuations
%favor the superconducting $s_{++}$-wave state without sign
%reversal, which is the extension of the
%previous charge quadrupole mechanism\cite{Kontani-Saito-Onari}.}

%%%%%%%%%%%%%%%%%%%%%%%%%

Below, we denote the five $d$-orbitals
$d_{3z^2-r^2}$, $d_{xz}$, $d_{yz}$, $d_{xy}$, and $d_{x^2-y^2}$ as
$l=1,2,3,4$, and $5$, respectively.
%%%%%%%%%%%%%%%%%%%%%%%%%%%
% Model and Hamiltonian
%%%%%%%%%%%%%%%%%%%%%%%%%%%
We analyze the following two-dimensional 
eight-orbital $d$-$p$ Hubbard model with parameter $r$
\cite{Onari-form}:
\begin{eqnarray}
H_{\rm M}(r)=H^0+rH^U, \ \ \
\label{eqn:Ham}
\end{eqnarray}
where 
$H^0$ is the unfolded tight-binding model for BaFe$_2$As$_2$ \cite{Nakaoka122},
FeSe \cite{Onari-form}, and NaFeAs; more details are presented  in Supplementary Material (SM) A \cite{SM}.
$H^U$ is the first-principles screened $d$-electron Coulomb potential in
each compound \cite{Arita}, and $r$ is the reduction parameter,
\color{black}which is
approximately proportional to the
renormalization factor $z$ in the coherence part of the Green function
\cite{FeSe-Yamakawa}. We note that $r$ is the unique free parameter in the present theory.
We set $r$ to reproduce experimental weak (FeSe) or moderate (NaFeAs, BaFe$_2$As$_2$) spin fluctuation strength in the RPA.
\color{black}

%%%%%%%%%%%%%%%%%%%%%%%%%%%%%%%%%
%%%%%%%%%%%%%%%%%%%%%%%%%%%%%%%%%

First, we focus on the unfolded BaFe$_2$As$_2$ model directly given by
 the first-principles calculation using WIEN2k.
 Figure \ref{fig:FS}(b) shows the unfolded Fermi surfaces
(FSs).
The size of h-FS3 around M point composed of orbital 4 is
similar to that of e-FS1(2) around X(Y) point, which results in a good
inter-orbital nesting.
We calculate the spin (charge) susceptibilities ${\hat \chi}^{s(c)}(q)$
for $q=(\q,\w_m=2m\pi T)$ based on the random-phase-approximation (RPA).
The spin Stoner factor $\alpha_{s}$ 
is defined as
the maximum eigenvalue of $\hat{\Gamma}^{s}\hat{\chi}^0(\bm{q},0)$, where
${\hat \Gamma}^{s(c)}$ is the bare Coulomb interaction 
for the spin (charge) channel,
and $\hat{\chi}^0$ is the irreducible susceptibilities
given by the Green function without self-energy 
${\hat G}(k)=[(i\e_n-\mu){\hat1}-{\hat{h}}^0(\k)]^{-1}$ 
for $=[\k,\e_n=(2n+1)\pi T]$. Since the relation
 $\hat{\chi}^s(\q)\propto\frac{1}{1-\a_s}$ holds, spin fluctuations
 become large with increasing $\alpha_s$ $(\propto r)$, and
 $\a_s=1$ corresponds to
spin-ordered state. 
Here, ${\hat{h}}^0(\k)$ is the matrix expression of $H^0$ 
and $\mu$ is the chemical potential.
Details of $\hat{\Gamma}^{s(c)}$, $\hat{\chi}^{s(c)}(q)$, 
and $\hat{\chi}^0(q)$ are presented in SM A \cite{SM}.
We fix the parameters $r=0.303$ in the BaFe$_2$As$_2$ model unless
 otherwise noted. In this case, $\a_s=0.96$ at
 $T=30$meV, and the averaged intra-orbital Coulomb
 interaction is $rU\sim 1.6$eV. 
Figure \ref{fig:FS}(c) shows the obtained spin susceptibilities
$\chi^{s}_{4,4;4,4}(\q,0)$ and $\chi^{s}_{3,4;3,4}(\q,0)$,
the peaks of which at $\q=(0,\pi)$ originate from the
intra-orbital (4-4) and the inter-orbital (3-4) nesting, respectively. 
$\chi^{s}_{4,4;4,4}$ is larger than $\chi^{s}_{3,3;3,3}$ because of the good
intra-$d_{xy}$-orbital nesting between e-FSs and h-FS3.

Hereafter, we study the symmetry breaking in the self-energy 
$\hat{f}^\q$ for wavevector $\q$ based on the density-wave (DW) equation introduced in 
Ref. \cite{Onari-form,Kawaguchi-Cu,Onari-B2g}. 
We calculate both the momentum and orbital dependences of $\hat{f}$
self-consistently 
to analyze both the orbital order and bond order on equal footing.
To identify the realized DW with wavevector $\q$,
we solve the linearized DW equation:
%self-consistent equations with respect to the symmetry-breaking
%self-energy $\Delta\hat{\Sigma}^{\q}$, and obtain the linearized CDW equation:
%
\begin{eqnarray}
&& \lambda_\q f^\q_{l,l'}(k)= \frac{T}{N}
\sum_{k',m,m'} {K}^{\bm{q}}_{l,l';m,m'}(k,k')f^\q_{m,m'}(k'),
\label{eqn:linearized}    
\end{eqnarray}
where $\lambda_\q$ is the eigenvalue of the DW equation.
%Feynman diagram of the linearized CDW equation (\ref{eqn:linearized}) is depicted in Fig. \ref{fig:FS} (e).
The DW with wavevector $\q$ appears when $\lambda_\q=1$, and
the eigenvector $\hat{f}^\q(k)$ gives the DW form factor.
A larger value of $\lambda_\q$ corresponds to a more dominant DW state.
Details of the kernel function
$\hat{K}^{\bm{q}}(k,k')$ are given in SM A \cite{SM}.
The Maki--Thompson (MT) terms and Aslamazov--Larkin (AL) terms shown in
Fig. \ref{fig:FS}(d) are included in the kernel function.
Near the magnetic criticality,
the AL terms are strongly enhanced in proportion to
$\sum_p\chi^s(p)\chi^s(-p+q_c)$, and they induce charge DW order through the three-boson coupling $C_{\Q_s,\Q'_s}$ in Fig. \ref{fig:FS}(a).

%In this study, we require Hermicity of the symmetry-breaking self-energy, 
%and neglect the frequency dependence of it as follows:
%$\Delta\hat{\Sigma}^\q(\k,\e_n>0)=0.5\left[\Delta\hat{\Sigma}^\q(\k,0+i\delta)+%\Delta\hat{\Sigma}^\q(\k,0-i\delta)\right]\equiv\Delta\hat{\Sigma}^\q(\k)$,
% and $f^\q_{l,m}(\k,\e_n<0)=\left[f^\q_{m,l}(\k)\right]^*$.

%%%%%%%%%%%%%%%%%%%%%%%%%%%%%%%%%
\begin{figure}[!htb]
\includegraphics[width=.9\linewidth]{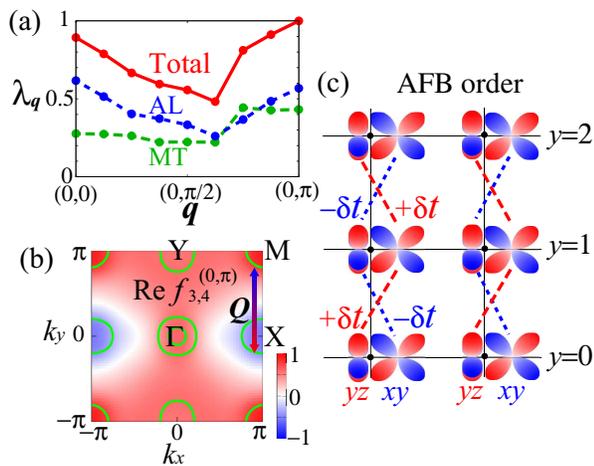}
\caption{
(a) Obtained $\q$ dependences of $\lambda_{\q}$ at $T=32.4$meV. The
 contributions from the AL and MT terms are also shown.
(b) Dominant component of the form factor at $\q=(0,\pi)$, $f_{3,4}^{\q}(\k)$,
 which is given by the
 off-diagonal orbitals 3 and 4.
The green lines indicate FSs. 
(c) Picture of the inter-orbital AFB order
  induced by $f_{3,4}^{(0,\pi)}$.
}
\label{fig:Sigma}
\end{figure}
%%%%%%%%%%%%%%%%%%%%%%%%%%%%%%%%%

%From the solution ${\hat f}^\q(k)$ of the DW equation, we
%derive the static form factor ${\hat f}^\q(\k)$ at $\e=0$ based on 
%analytic continuation.
Figure \ref{fig:Sigma}(a) shows the $\q$-dependences of $\lambda_\q$ for
the total terms, the MT
terms, and AL terms at $T=32.4$meV.
$\q=(0,\pi)$ AFB order appears at $T^*=32.4$meV, while
$\lambda_{\bm{0}}$ is slightly smaller than unity. 
Thus, the ferro-orbital transition temperature $T_S$ is lower than
$T^*$. The relation
$\lambda_{(0,\pi)}>\lambda_{\bm{0}}$ is robust in the presence of
moderate spin fluctuations $\a_s\gtrsim 0.85$.
Both the AL and MT terms contribute to the AFB order cooperatively as shown in Fig. \ref{fig:Sigma}(a). 
Figure \ref{fig:Sigma}(b) shows
the dominant component of the static form factor, $f^{\q}_{3,4}(\k)$, for
$\q=(0,\pi)$, which is derived from the analytic continuation of $\hat{f}^{\q}(k)$.
The other sub-dominant components are
explained in SM B \cite{SM}.
Focusing on the X and M points, $f^{(0,\pi)}_{3,4}(\k)$ is proportional to
$-\cos(k_y)$, which corresponds to the inter-orbital AFB order shown in
Fig. \ref{fig:Sigma}(c),
where the $y$-direction hoppings between orbitals 3 and 4 are
modulated by the correlation hopping $\delta t_{3,4}(y;y\pm1)=-\delta
t_{4,3}(y;y\pm1)=\delta t(-1)^y$. Note that $\delta t_{l,m}(y;y')$ is
real and equal to $\delta
t_{m,l}(y';y)$.

The origin of the AFB order $f^{(0,\pi)}_{3,4}$ is the quantum
interference between the antiferro-spin fluctuations
$\chi^s_{3,4;3,4}(\Q)$ for $\Q\approx (0,\pi)$ and ferro-spin fluctuations $\chi^s_{4,4;4,4}(\bm{0})$
shown in Fig. \ref{fig:FS}(a).
The former is enhanced by
 the inter-orbital nesting shown in Fig. \ref{fig:FS}(b), while the latter $\chi^s_{4,4;4,4}(\bm{0})$ is
caused moderately by the forward intra-orbital scattering of
orbital 4 in h-FS3.
The developments of $\chi^s_{3,4;3,4}(\Q)$ and $\chi^s_{4,4;4,4}(\bm{0})$ are shown by the red
dotted circle in Fig. \ref{fig:FS}(c).
Moreover, the three-boson coupling $C_{\Q_s,\Q'_s}$ in Fig. \ref{fig:FS}(a) is strongly
enlarged when $\q_c=\Q_s+\Q'_s$ is a nesting vector \cite{Yamakawa-Cu},
and this condition is satisfied when $\Q_s=\Q$ and $\Q'_s=\bm{0}$. Thus, $\lambda_{\Q}$ becomes large due to the AL terms.
In addition to the AL terms, the MT terms 
%given by $\chi^s_{3,3;4,4}(\Q)$
strengthen the sign change of
$f^{(0,\pi)}_{3,4}(\k)$ between X and M points, as reported previously \cite{Onari-form,Onari-B2g,Chubukov-FeSe}. Thus, the 
AFB order originates from the cooperation between the AL and
MT terms due to the inter-orbital nesting.

In contrast, the FO instability that corresponds to $\lambda_{\bm{0}}$
originates mostly from the AL term owing to the combination of
$\chi^s_{3,3;3,3}(\q)$ and $\chi^s_{3,3;3,3}(-\q)$ for $\q\approx(\pi,0)$, as discussed in Refs. \cite{Onari-SCVC,Onari-SCVCS,FeSe-Yamakawa}.

%%%%%%%%%%%%%%%%%%%%%%%%%%%%%%%%%
\begin{figure}[!htb]
\includegraphics[width=.8\linewidth]{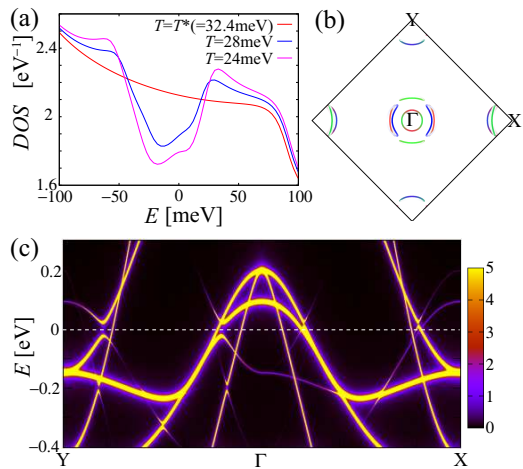}
\caption{
(a) DOS at $T=T^*(=32.4$meV), $28$meV, and $24$meV
 in the AFB state. Here, we introduced the quasiparticle damping
$\gamma=10$meV.
(b) FSs and (c) spectral weight for $\gamma=1$meV under $\q=(0,\pi)$ AFB order at
 $T=28$meV in the original two-Fe Brillouin zone. 
}
\label{fig:folded}
\end{figure}
%%%%%%%%%%%%%%%%%%%%%%%%%%%%%%%%%

Here, we explain that a pseudogap originates from the band-folding
 driven band-hybridization under the $\q=(0,\pi)$ AFB order.
 Since the form factor grows in proportion to Re$\sqrt{\lambda_\q-1}$ in the Ginzburg-Landau theory, we introduce the
mean-field-like $T$-dependent form factor
$\hat{f}^\q(T)=f^{\rm
max}\tanh\left(1.74\sqrt{T^*/T-1}\right)\hat{f}^\q_{\rm DW}$, where
$\hat{f}^\q_{\rm DW}$ is the obtained form factor normalized as
$\max_{\k} |f^\q_{\rm DW}(\k)|=1$.
We put $f^{\rm max}=60$meV. Details of calculation method under the AFB order are explained in SM B \cite{SM}.
Figure \ref{fig:folded}(a) shows the obtained DOS.
For $T<T^*$, a pseudogap appears, which is consistent with the
ARPES measurement \cite{Shimojima-PG}.
Figures \ref{fig:folded}(b) and \ref{fig:folded}(c) show the FSs and
spectral weight, respectively under the $\q=(0,\pi)$ AFB order at
 $T=28$meV.
Here, the folded band structure under the AFB order is unfolded to the
original two-Fe Brillouin zone by following Ref. \cite{Wei-Ku},
 which gives the spectrum corresponding to the
 ARPES measurements \cite{Shimojima-PG,Shimojima-SB,Fujimori-PV}.
Owing to the band-folding, several Dirac-type bandstructures and shadow bands
appear, as reported through an ARPES study \cite{Fujimori-PV}.

%We note that the orbital component of form
%factor is significantly different between the AFB order and the FO order.
%Note that the $\q=(0,\pi)$ AFO state in the present unfolded
%eight-orbital model is identical to the $\q=(\pi,0)$ AFO state in the
%folded sixteen-orbital model\cite{Kontani-Saito-Onari}.

\begin{figure}[htb]
\includegraphics[width=.99\linewidth]{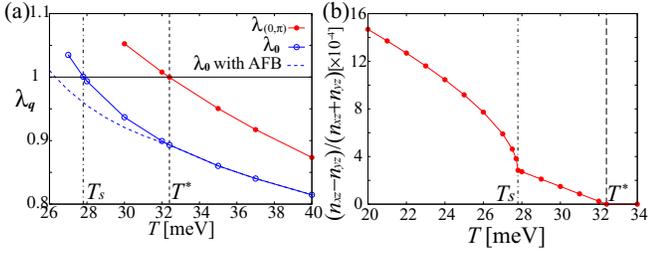}
\caption{
%(color online)
(a) $T$ dependencies of $\lambda_{\q}$ for $\bm{q}=(0,\pi)$ and $\bm{q}=\bm{0}$. The blue dotted line shows
 $\lambda_{\bm{0}}$ with the AFB order for $T<T^*$.
(b) The nematicity $\psi=(n_2-n_3)/(n_2+n_3)$ including both
  AFB order for $T<T^*$ and FO order for $T<T_S$.
}
\label{fig:T-dep}
\end{figure}

%which corresponds to the intersections of the
%orange dotted lines in Fig. \ref{fig:Kernel} (d),
In the following, we explain the ``hidden nature'' of the present AFB
order, that is, the tiny anomalies in  $\chi^s(\Q)$ and $\chi_{\rm
nem}(\bm{0})$ at $T^*$. This is a long-standing mystery in Ba122.
The $T$ dependences of 
$\a_s$ with and without
$\hat{f}^{(0,\pi)}(T)$ are shown in Fig. S2 in SM B \cite{SM}. The
AFB order suppresses $\a_s$ only slightly since the spin fluctuations
are essentially intra-orbital, while intra-orbital components
of $\hat{f}^{(0,\pi)}(T)$ are sub-dominant.
Next, we analyze the $T$ dependencies of eigenvalue
$\lambda_{\q}$ for 
the FO order and AFB order by following SM B \cite{SM}. As shown in Fig. \ref{fig:T-dep} (a), the
FO-order eigenvalue $\lambda_{\bm{0}}$ is suppressed only slightly by
the finite AFB order, 
owing to the slight decrease of $\a_s$ and the ``orbital selectivity'' in nematicity:
We stress that the dominant component of form factor is different between the
off-diagonal $f^{(0,\pi)}_{3,4}$ in the AFB order and the diagonal
$f^{\bm{0}}_{3,3(4,4)}$ ($f^{\bm{0}}_{3,4}=0$) in the FO order as shown in SM C \cite{SM}.
Thus, neither $\chi^s(\Q)\propto 1/(1-\a_s)$ nor $\chi_{\rm
nem}(\bm{0})\propto 1/(1-\lambda_{\bm{0}})$
 would show a visible
anomaly at $T^*$, which is consistent with experiments. (Here, the increment of $T_S$ due to the electron-phonon interaction is not considered for simplicity.)
In contrast, the FO order at $T_S$ causes a sizable anomaly for
$\chi^s(\Q)$ and $\chi_{\rm
nem}(\bm{0})$.

%%%%%%%%%%%%%%%%%%%%%%%%%%%%%%%%%
%%%%%%%%%%%%%%%%%%%%%%%%%%%%%%%%%

Another long-standing mystery is the $T$-linear behavior of nematicity $\psi$ in Ba122
\cite{Kasahara-torque} and NaFeAs \cite{NaFeAs-T*} below $T^*$. In order to solve this mystery, we calculate
 the $T$ dependence of nematicity $\psi=(n_2-n_3)/(n_2+n_3)$ in Fig. \ref{fig:T-dep}(b), where
both $\hat{f}^{(0,\pi)}(T)$ for $T<T^*$ and
the FO order $\hat{f}^{\bm{0}}(T)$ 
for $T<T_S$ are introduced.
For $T<T_S$, we assume
$\hat{f}^{\bm{0}}(T)=f^{\rm
max}\tanh\left(1.74\sqrt{T_S/T-1}\right)\hat{f}^{\bm{0}}_{\rm DW}$,
where $\hat{f}^{\bm{0}}_{\rm DW}$ is the 
obtained form factor normalized as $\max_{\k}|f_{\rm DW}^{\bm{0}}(\k)|=1$.
Details of $\hat{f}^{\bm{0}}_{\rm DW}$ are presented
in SM C \cite{SM}.
We employ $f^{\rm max}=60$meV, which corresponds to the energy split
$\sim 60$meV in
the ARPES measurements \cite{ARPES}.
%The tiny $\psi$ for $T_S<T<T^*$ is due to $n_2\gtrsim n_{3}$ by the
%AFB order, and the large $\psi$ for $T<T_S$ is due to $n_2>n_{3}$
%caused by the FO order. 
Because the AFB order only slightly suppresses the FO-order
transition as shown in Fig. \ref{fig:T-dep}(a), the obtained
multistage nematic transitions are naturally expected in Ba122.
The $T$-linear behavior $\psi\propto(T^*-T)$ for $T_S<T<T^*$ is a consequence of the relation
$\psi\propto[f^{(0,\pi)}(T)]^2$ because the $f^{(0,\pi)}$ term cannot
contribute to any $\q=\bm{0}$ linear response.
%the uniform $(\q=\bm{0})$ orbital polarization.
%Therefore, $\psi$ is independent of the sign of $f^{(0,\pi)}$. 
Note that the form factor $\hat{f}^{(\pi,0)}$ for
$\q=(\pi,0)$ gives $\psi<0$.
Thus, the $T$-linear behavior of $\psi$ below $T^*$ is also naturally
explained by the AFB order.
On the other hand, $\psi\propto\sqrt{T_S-T}$ for $T<T_S$ is
induced by the FO order. To summarize, long-standing mysteries in the
hidden-nematic phase, such as tiny anomalies in $\chi^s(\Q)$ and $\chi_{\rm
nem}(\bm{0})$ at $T^*$ and a $T$-linear $\phi$ below $T^*$, are naturally explained in the present AFB-order scenario. 
%%%%%%%%%%%%%%%%%%%%%%%%%%%%%%%%%
\begin{figure}[!htb]
\includegraphics[width=.85\linewidth]{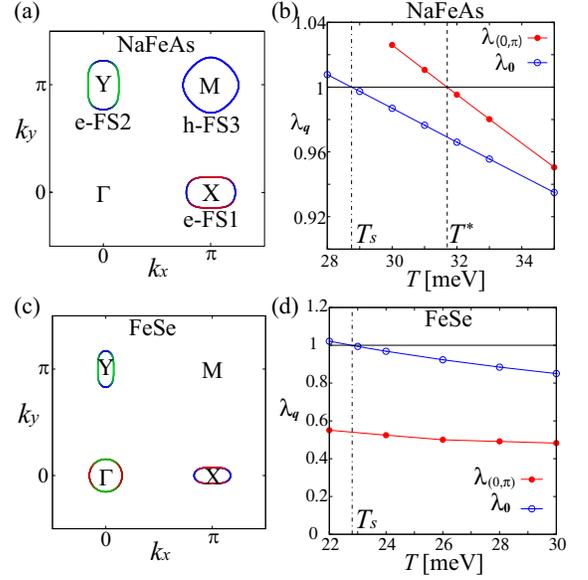}
\caption{
(a) FSs of the NaFeAs model in the unfolded zone.
(b) $\lambda_{\q}$ for $\bm{q}=(0,\pi)$ and $\bm{q}=\bm{0}$
 in the NaFeAs model for $r=0.339$ ($rU\sim1.8$eV). Then, $\a_s=0.92$ at
 $T=30$meV. 
(c) FSs of the FeSe model.
(d) $\lambda_{\q}$ in the FeSe model for $r=0.239$
 ($rU\sim1.7$eV). Then, $\a_s=0.85$ at
 $T=30$meV. }
\label{fig:nematic}
\end{figure}
%%%%%%%%%%%%%%%%%%%%%%%%%%%%%%%%%

Finally, we discuss the universality of the hidden nematic order by
focusing on NaFeAs and FeSe. 
%{\color{red}We impose the relation $\chi^s_{\rm
%BaFe_2As_2}(\Q)\gg\chi^s_{\rm NaFeAs}(\Q)>\chi^s_{\rm FeSe}(\Q)$, which
%is consistent with the NMR results.} 
According to the ARPES measurement in NaFeAs \cite{NaFeAs-ARPES,NaFeAs-ARPES2}, 
only a single hole-band mainly composed of $d_{xy}$-orbital crosses the
Fermi level, resulting from the spin-orbit interaction (SOI)-induced
band hybridization.
To reproduce the single $d_{xy}$-orbital-like hole-pocket in  NaFeAs, we introduce
the NaFeAs model without SOI by shifting downwards the $d_{xz}$ and
$d_{yz}$ hole-bands immediately below the Fermi level in the BaFe$_2$As$_2$
model.
The FSs in the NaFeAs model are shown in
Fig. \ref{fig:nematic} (a).
Details of the model are presented in SM A \cite{SM}.
The obtained $T$ dependences of
$\lambda_{\q}$ in Fig. \ref{fig:nematic} (b) is similar to that in the
BaFe$_2$As$_2$ model.
The AFB order in NaFeAs is naturally understood as a consequence of
the inter-orbital nesting between h-FS3 and e-FSs, as we discussed
above. The FO order is driven by the spin fluctuations on the $d_{xz}$ and $d_{yz}$
orbitals. They are not weak because the top of the $d_{xz}$ and
$d_{yz}$-orbital hole-band in NaFeAs is very close to the
Fermi level according to the ARPES
measurements \cite{NaFeAs-ARPES,NaFeAs-ARPES2}.
The derived multistage nematic transition is consistent with the experiment
on NaFeAs \cite{NaFeAs-T*}.
On the other hand, the $d_{xy}$-orbital hole pocket is missing in the FeSe model shown in
Fig. \ref{fig:nematic} (c).
Because the inter-orbital nesting is missing, $\lambda_{(0,\pi)}$ in
the FeSe model is considerably smaller than unity as shown in
Figs. \ref{fig:nematic} (d) and S4 (b) in SM D \cite{SM}, which is consistent with the absence of the hidden nematic order in
FeSe \cite{Shimojima-FeSe}.

In this numerical study, we neglected the self-energy. However, the
results are essentially unchanged if the self-energy is incorporated
into the DW equation, as we verified in SM E \cite{SM}:
We incorporate the self-energy into the DW equation in the framework of the conserving approximation, where the macroscopic conservation laws are
satisfied rigorously and unphysical results are avoided. 

In summary, 
we demonstrated that the origin of the hidden nematic state for $T_S<T<T^*$ in
 BaFe$_2$As$_2$ and NaFeAs, which is a long-standing unsolved problem,
 is naturally explained as the AFB ordered state.
The tiny $T$-linear nematicity $\psi=(n_{xz}-n_{yz})/(n_{xz}+n_{yz})$
as well as the emergence
of the pseudogap and shadow band are naturally explained based on the
present scenario.
The phase diagrams of Ba122 and NaFeAs are understood by the present multistage nematic
transition scenario.
In contrast, the hidden nematic order is absent in FeSe because of the absence of the $d_{xy}$-orbital hole pocket. 

Finally, we stress that the bond fluctuations significantly
contribute to the pairing mechanism, as explained
 in SM F \cite{SM}. We will discuss this novel pairing mechanism in detail in future publications.

%%%%%%%%%%%%%%%%%%%%%
\acknowledgements
We are grateful to 
%Y. Matsuda, T. Shibauchi, A. Fujimori, S. Shin, and
 Y. Yamakawa
for useful discussions.
This work was supported
by Grants-in-Aid for Scientific Research from MEXT,
Japan (No. JP19H05825, JP18H01175, and JP17K05543)

%%%%%%%%%%%%%%%%%%%%%%%
%\appendix
%\section{Supplemental Material}

%%%%%%%%%%%%%%%%%%%%%%%%
%references
%%%%%%%%%%%%%%%%%%%%%%%%

%\end{document}
%%%%%%%%%%%%%%%%%%%%%%%%%%%%%%%%%%%%%%%
\clearpage

\makeatletter
\renewcommand{\thefigure}{S\arabic{figure}}
\renewcommand{\theequation}{S\arabic{equation}}
\makeatother
\setcounter{figure}{0}
\setcounter{equation}{0}
\setcounter{page}{1}
\setcounter{section}{1}

\begin{widetext}
\begin{center}
{\bf 
[Supplementary Material] \\
Hidden antiferro-nematic order in 
Fe-based superconductor BaFe$_2$As$_2$ and NaFeAs
above $T_S$
}%
\end{center}

\begin{center}
Seiichiro Onari and Hiroshi Kontani
\end{center}

\begin{center}
\textit{Department of Physics, Nagoya University, Nagoya 464-8602, Japan}
\end{center}

\end{widetext}
\subsection{A: Eight-orbital models for BaFe$_2$As$_2$ and FeSe}
Here, we introduce the eight-orbital $d$-$p$ models for BaFe$_2$As$_2$,
 FeSe, and NaFeAs that are analyzed in the main text.
We first derive first-principles tight-binding models
by using the WIEN2k \cite{WIEN2k} and WANNIER90 \cite{wannier90} codes.
The model for BaFe$_2$As$_2$ is directly given by the first-principles
 tight-binding model without any modification.

Next, to obtain the experimentally observed Fermi surfaces
(FSs) in FeSe,
we introduce the $k$-dependent shifts for orbital $l$, $\delta E_l$,
by introducing the intra-orbital hopping parameters,
as explained in Ref. \cite{S-FeSe-Yamakawa}.
We shift the $d_{xy}$-orbital band [$d_{xz/yz}$-orbital band] 
at ($\Gamma$, M, X) points
by ($-0.6$eV, $-0.25$eV, $+0.24$eV) [($-0.24$eV, $0$eV, $+0.12$eV)] for FeSe.
\color{black}The NaFeAs model is constructed by shifting the
 $d_{xz/yz}$-orbital band of the BaFe$_2$As$_2$ model at $\Gamma$ point
 by $-0.3$eV in order to reproduce the experimental FSs of NaFeAs
 without the SOI.
\color{black}

We also introduce the mass-enhancement factor $z^{-1}=1.6$
 \color{black} \color{black} for the $d_{xy}$
orbital in FeSe, \color{black}while $z^{-1}=1$ for other
 orbitals.
\color{black}

\begin{figure}[!htb]
\includegraphics[width=.99\linewidth]{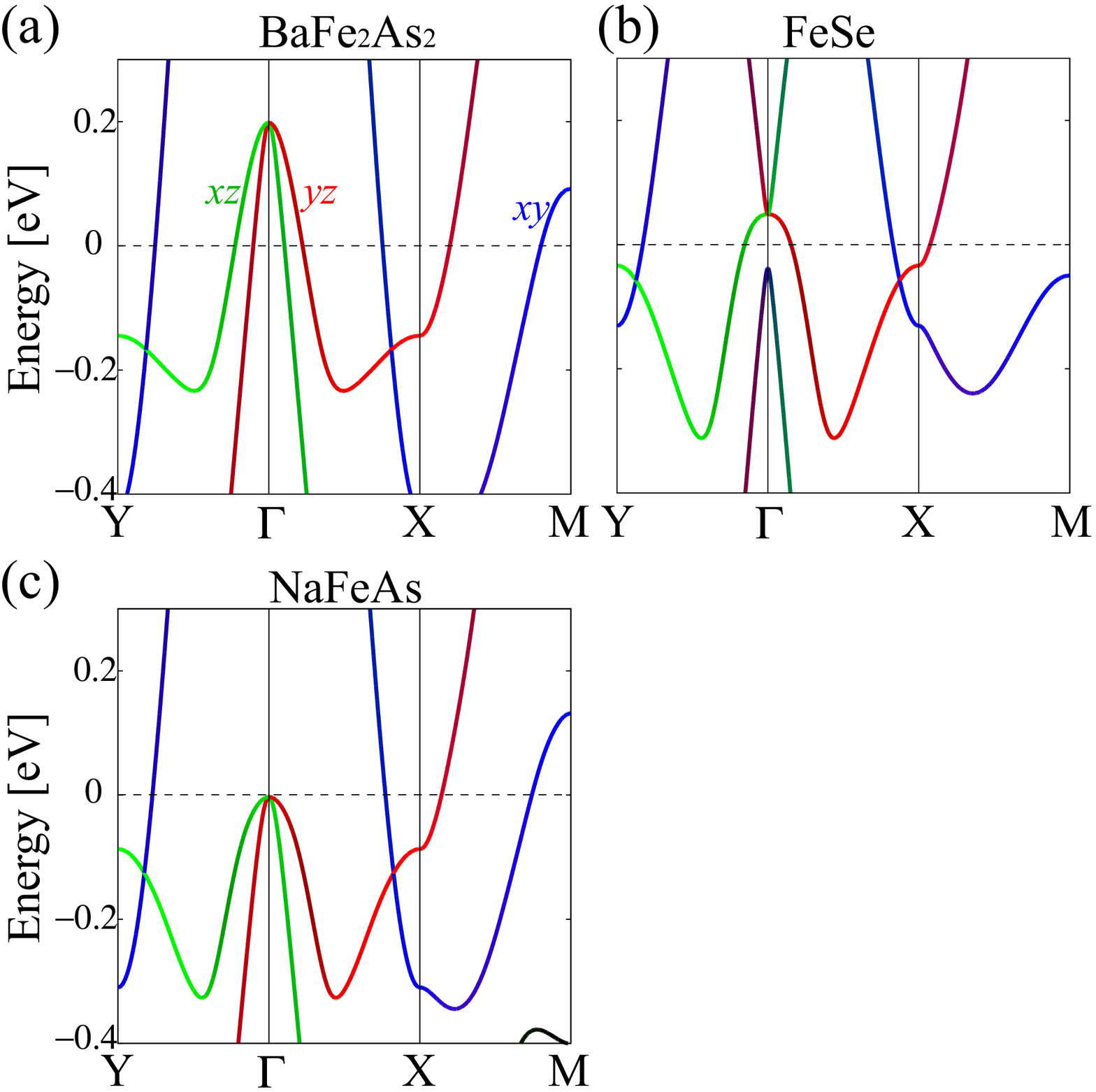}
\caption{
Obtained bandstructures of (a) the BaFe$_2$As$_2$ model, (b) FeSe model,
 \color{black} and (c)
 NaFeAs model. \color{black}
}
\label{fig:band}
\end{figure}

Figure \ref{fig:band} shows the bandstructures of 
the obtained BaFe$_2$As$_2$ model, FeSe model, \color{black} and NaFeAs model.\color{black}

%%%%%%%%%%%%%%%%%%%%%%%%%%%%%%%%%
%%%%%%%%%%%%%%%%%%%%%%%%%%%%%%%%%

Here, we explain the $d$-orbital Coulomb interaction introduced by the constraint RPA
(cRPA) method for each compound in Ref. \cite{S-Arita}.
The Coulomb interaction for the spin channel 
in the main text is
\begin{equation}
(\Gamma^{\mathrm{s}})_{l_{1}l_{2},l_{3}l_{4}} = \begin{cases}
U_{l_1,l_1}, & l_1=l_2=l_3=l_4 \\
U_{l_1,l_2}' , & l_1=l_3 \neq l_2=l_4 \\
J_{l_1,l_3}, & l_1=l_2 \neq l_3=l_4 \\
J_{l_1,l_2}, & l_1=l_4 \neq l_2=l_3 \\
0 , & \mathrm{otherwise}.
\end{cases}
\end{equation}
Furthermore, the Coulomb interaction for the charge channel is
\begin{equation}
({\hat \Gamma}^{\mathrm{c}})_{l_{1}l_{2},l_{3}l_{4}} = \begin{cases}
-U_{l_1,l_1}, & l_1=l_2=l_3=l_4 \\
U_{l_1,l_2}'-2J_{1_1,l_2} , & l_1=l_3 \neq l_2=l_4 \\
-2U_{l_1,l_3}' + J_{l_1,l_3} , & l_1=l_2 \neq l_3=l_4 \\
-J_{1_1,l_2} , &l_1=l_4 \neq l_2=l_3 \\
0 . & \mathrm{otherwise}.
\end{cases}
\end{equation}

The averaged intra-orbital Coulomb interaction is 5.2eV and 7.2eV
for BaFe$_2$As$_2$ and FeSe compounds, respectively \cite{S-Arita}.

By using the multiorbital Coulomb interaction,
the spin (charge) susceptibility in the RPA is given by  
\begin{equation}
{\hat \chi}^{s(c)}(q)={\hat\chi^0}(q)[1-{\hat \Gamma}^{s(c)}{\hat
\chi^0(q)}]^{-1},
\end{equation}
where the irreducible susceptibility is
\begin{equation}
\chi^0_{l,l';m,m'}(q)= -\frac{T}{N}\sum_k
G_{l,m}(k+q)G_{m',l'}(k).
\end{equation}
Here, ${\hat G}(k)$ is the multiorbital Green function 
introduced in the main text.

The kernel function
$\hat{K}^{\bm{q}}(k,k')$ \cite{S-Onari-form,S-Kawaguchi-Cu} is given by
\begin{equation}
K^{\bm{q}}_{l,l';m,m'}(k,k')=-\sum_{m_1,m_2}I^{\bm{q}}_{l,l';m_1,m_2}(k,k')g^{\bm{q}}_{m_1,m_2;m,m'}(k'),
\label{eqn:K} 
\end{equation}
where 
$g^{\bm{q}}_{l,l';m,m'}(k)\equiv
G_{l,m}\left(k+\bm{q}\right)G_{m',l'}(k)$,
and $\hat{I}^{\bm{q}}(k,k')$ is the four-point vertex.

$\hat{I}^{\bm{q}}(k,k')$ is given as
\begin{eqnarray}
&& \!\!\!\!\!\!\!\!\!\!\!
I^{\bm{q}}_{l,l';m,m'}(k,k')=\sum_{b=s,c}
\left[-\frac{a^b}{2} V^{b}_{l,m;l',m'}(k-k')\right.
\nonumber \\
&& 
+\frac{T}{N}\!\!\!\!\sum_{p,l_1,l_2,m_1,m_2}\!\!\!\!\!\!\!\!\!\!
 \frac{a^b}{2} V^{b}_{l,l_1;m,m_2}\left(p+{\q}\right)V^{b}_{m',l_2;l',m_1}\left(p\right)
 \nonumber \\
&& \qquad\qquad
\times G_{l_1,m_1}(k-p)G_{l_2,m_2}(k'-p)
\nonumber \\
&&
+\frac{T}{N}\!\!\!\!\sum_{p,l_1,l_2,m_1,m_2}\!\!\!\!\!\!\!\!\!\!
 \frac{a^b}{2} V^{b}_{l,l_1;l_2,m'}\left(p+\q\right)V^{b}_{m_2,m;l',m_1}\left(p\right)
 \nonumber \\
&& \qquad\qquad
\left.\times G_{l_1,m_1}(k-p)G_{l_2,m_2}(k'+p+\q)\right],
%-({\rm Double\;counting\;} [\hat{\Gamma}^{s(c)}]^2 \;{\rm terms})
\label{eqn:S-K} 
\end{eqnarray}
%\begin{eqnarray}
%&& \!\!\!\!\!\!\!\!\!\!\!
%K^{\bm{q}}_{l,l';m,m'}(k,k')=
%\sum_{l'',m''}\left[\frac32 V^{s}(k-k')+\frac12 V^{c}(k-k')\right]_{l,l'';l',m''}
%\nonumber \\
%&& \ \ \ \ \ \ \ \ \ \ 
%\times G^0_{m',m''}\left(k'-\frac{\q}{2}\right)G^0_{l'',m}\left(k'+\frac{\q}{2}\right)
% \nonumber \\
%&&
%-\frac{T}{N}\!\!\!\!\sum_{p,l_1,l_2,l_3,m_1,m_2,m_3}\!\!\!\!\!\!\!\!\!\!\!\!\!\!
% \left[ \frac32 V^{s}_{l_1,l_2;l',m_1}\left(p-\frac{\q}{2}\right)V^{s}_{l,l_3;m_2,m_3}\left(p+\frac{\q}{2}\right) \right.
% \nonumber \\
%&& 
%\left. +\frac12 V^{c}_{l_1,l_2;l',m_1}\left(p-\frac{\q}{2}\right)V^{c}_{l,l_3;m_2,m_3}\left(p+\frac{\q}{2}\right) \right]G^0_{l_3,m_1}(k-p)
%\nonumber \\
%&& 
%\times \left[{\Lambda}^\q_{m',l_1;l_2,m_3;m_2,m}(k';p)+{\Lambda}^\q_{m',m_3;m_2,l_1;l_2,m}(k';-p)\right],
%%-({\rm Double\;counting\;} [\hat{\Gamma}^{s(c)}]^2 \;{\rm terms})
%\label{eqn:S-K} 
%\end{eqnarray}
%
where $a^s=3$, $a^c=1$, $p=(\p,\w_l)$, and
$\hat{V}^{s(c)}(q)=\hat{\Gamma}^{s(c)}+\hat{\Gamma}^{s(c)}\hat{\chi}^{s(c)}(q)\hat{\Gamma}^{s(c)}$.
%in Ref. \cite{Onari-SCVC,Onari-SCVCS}.

In Eq. (\ref{eqn:S-K}),
the first line corresponds to the Maki-Thompson (MT) term,
and the second and third lines give the AL1 and AL2 terms, respectively.
In the MT term,
the first-order term with respect to ${\hat{\Gamma}}^{s,c}$ 
gives the Hartree--Fock (HF) term in the mean-field theory.

\subsection{B: Tiny anomaly of $\a_s$ at $T^*$}
\begin{figure}[!htb]
\includegraphics[width=.6\linewidth]{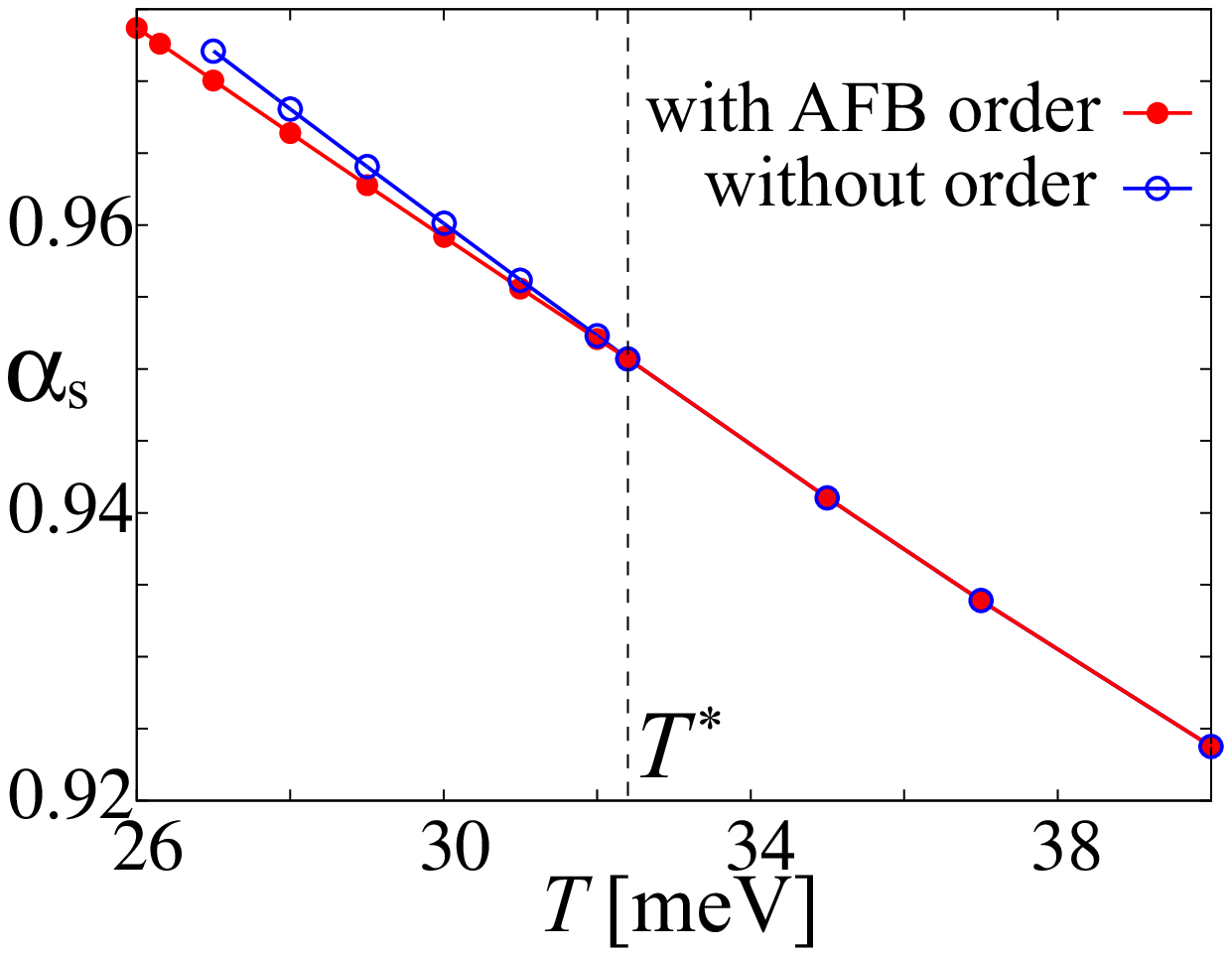}
\caption{
 $T$ dependencies of $\a_s$ with and without the AFB
 order.
}
\label{fig:alphas}
\end{figure}

Here, we analyze the $T$ dependence of 
$\a_s$ with and without AFB order
$\hat{f}^{(0,\pi)}(T)$.
The AFB order $\hat{f}^{(0,\pi)}(T)$ gives
modulation of the correlated hopping with period 2 along the $y$
direction shown in Fig. \ref{fig:Sigma} (c), which requires two times larger unit cell. By analyzing the
Hamiltonian in the two times larger unit cell, we obtain a pseudogap due
to the band-folding driven band-hybridization.
We calculate the spin susceptibilities in the two times larger unit
cell. 
As shown in Fig. \ref{fig:alphas},
 the AFB order suppresses $\a_s$ only slightly below $T^*$ since the
 diagonal component of form factor, which mainly affect the value of
 $\a_s$, is sub-dominant as mentioned below.
Thus, the tiny anomalies in $\chi^s(\Q)\propto \frac{1}{1-\a_s}$ at $T^*$, which is a long-standing mystery in Ba122, is
naturally explained by the AFB order.

By using the spin susceptibilities in the two times larger unit cell, we develop the method of DW equation under the
AFB order, which is complicated and heavy numerical calculation. 

In the main text, the dominant form factor is $f^{(0,\pi)}_{3,4}$ shown in Fig. \ref{fig:Sigma}(b).
$f^{(0,\pi)}_{4,4}$ and $f^{(0,\pi)}_{3,3}$ also emerge as the
sub-dominant form factors, which
 correspond to the intra-orbital AFB order with the modulation of correlation
hopping $\pm \delta t$  with period 2 along the $y$
direction.

The single-component GL 
theory considering terms up to the sixth order in
Ref. \cite{S-Kasahara-torque} leads to the prediction that the second-order transition at $T=T^*$
and the meta-nematic transition at $T=T_S$. In the present theory, in contrast, both
transitions are second-order owing to the multiorbital
component of the form factor.
In addition, the $T$-linear dependence of $\psi$ for $T_S<T<T^*$ is
naturally explained by the AFB order in our theory.

\color{black}
%%%%%%%%%%%%%%%%%%%%%%%%%%%%%%%%%
%%%%%%%%%%%%%%%%%%%%%%%%%%%%%%%%%
%%%%%%%%%%%%%%%%%%%%%%%%%%%%%%%%%
%\begin{figure}[!htb]
%\includegraphics[width=.8\linewidth]{figS3.eps}
%\caption{
% $T$ dependencies of $\a_s$ with and without the AFB
% order in the BaFe$_2$As$_2$ model.
%}
%\label{fig:Tdep2}
%\end{figure}

\subsection{C: Form factor of FO order}
\begin{figure}[!htb]
\includegraphics[width=.8\linewidth]{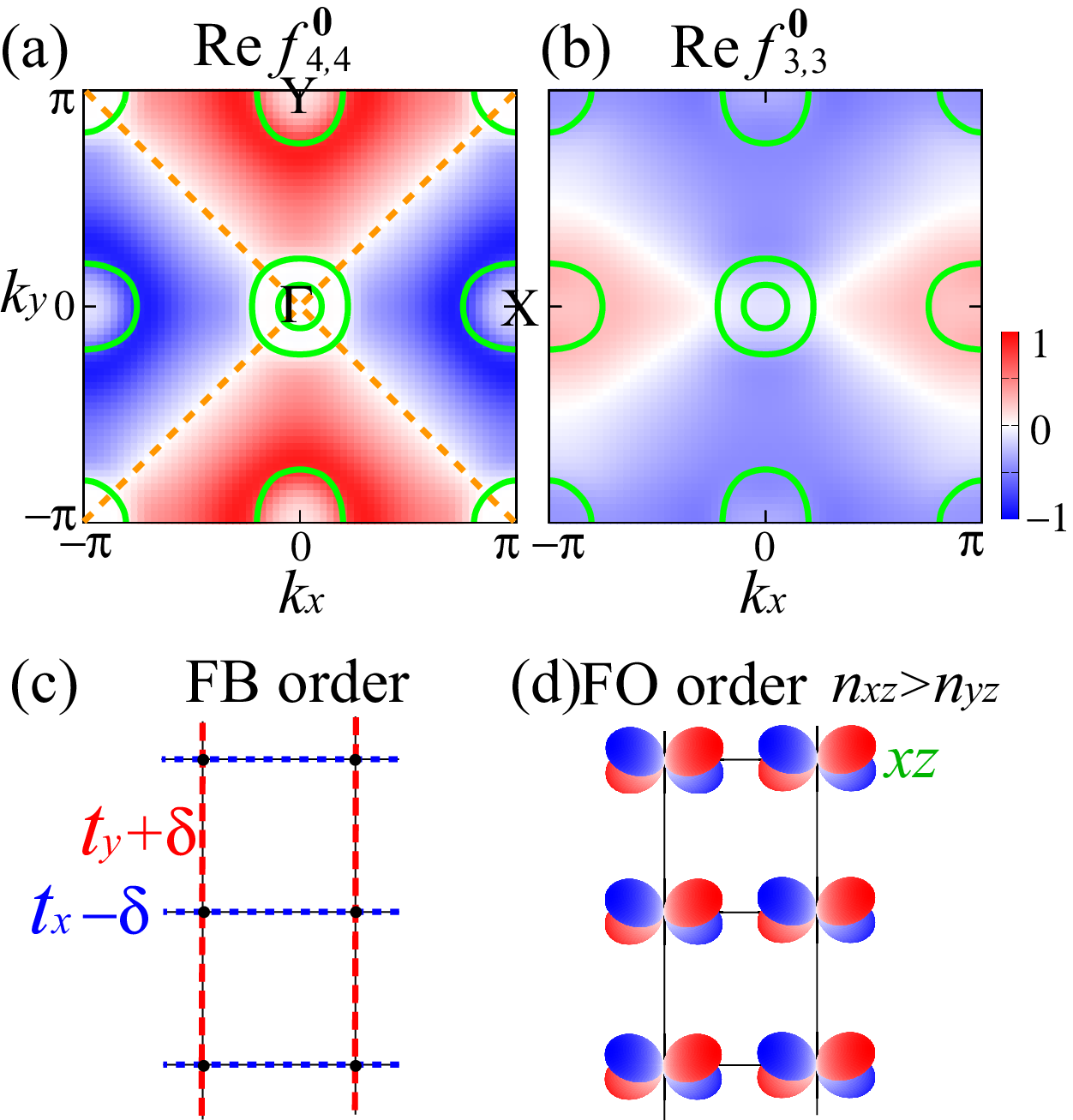}
\caption{
Form factors at $\q=\bm{0}$
obtained as the second-largest eigenvalue in the BaFe$_2$As$_2$ model on
(a) orbital 4 and (b) orbital 3. The green lines denote FSs without the self-energy.
The form factor for orbital 4, $f^{\bm{0}}_{4,4}\propto \cos k_x-\cos k_y$,
results in the ferro-bond (FB) order.
Schematic pictures of (c) FB order due to $f^{(0,\pi)}_{4,4}$ and 
(d) FO order due to $f^{\bm{0}}_{3,3(2,2)}$.
}
\label{fig:Sigma2}
\end{figure}

Here, we explain the obtained $\q=\bm{0}$ form factor in BaFe$_2$As$_2$.
Figures \ref{fig:Sigma2} (a) and (b) show the dominant form factors for
$\q=\bm{0}$.
$f^{\bm{0}}_{4,4}\propto \cos k_x-\cos k_y$ corresponds to the $B_{1g}$ bond order
of orbital 4 as shown in Fig. \ref{fig:Sigma2} (c). 
Figure \ref{fig:Sigma2} (b) shows 
$f^{\bm{0}}_{3,3}(k_x,k_y)= -f^{\bm{0}}_{2,2}(-k_y,k_x)$,
which causes the $B_{1g}$ FO order $n_{2}>n_{3}$ in
Fig. \ref{fig:Sigma2} (d). The off-diagonal $f^{\bm{0}}_{3,4}$ is $0$, while the $f^{(0,\pi)}_{3,4}$ is dominant in the AFB order.
Note that the $f^{\bm{0}}_{3,3}$ gives sign reversal orbital order (OO),
which is consistent with ARPES measurements in FeSe, Ba122 and La1111 \cite{Shimojima-FeSe2,Shen-reversal,La1111-ARPES}. 
\color{black}
%%%%%%%%%%%%%%%%%%%%%%%%%%%%%%%%%

\subsection{D: Results in FeSe model}
In the FeSe model employed in main text, we put the mass-enhancement
factor $z^{-1}=1.6$ for $d_{xy}$ orbital by following Ref. \cite{S-Onari-form}.
Here, we show that the qualitative results do not depend on the value of
$z$ by calculating the case of $z=1$ (without any mass-enhancement factor) in the FeSe model.

\begin{figure}[!htb]
\includegraphics[width=.9\linewidth]{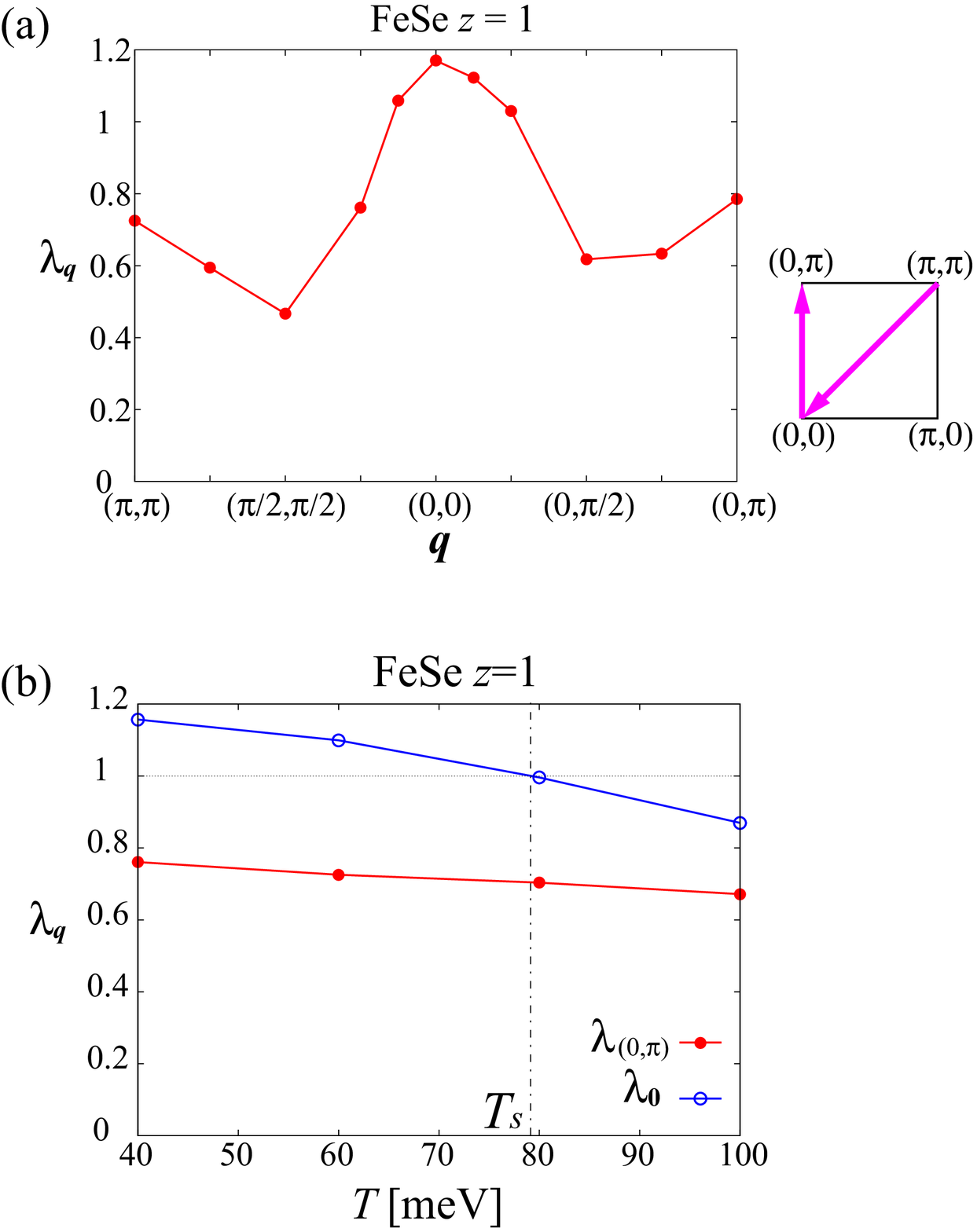}
\caption{
(a) Obtained $\q$ dependence of $\lambda_{\q}$ at $T=30$meV
 in the FeSe model for $z=1$ and $r=0.221$ ($rU\sim1.6$eV). Then, $\a_s=0.85$ at
 $T=30$meV. The $\q$ path is shown by arrows in the right panel.
(b) $T$-dependences of $\lambda_{\q}$ for $\bm{q}=(0,\pi)$ and $\bm{q}=\bm{0}$
 in the FeSe model for $z=1$ and $r=0.221$. 
}
\label{fig:z=1}
\end{figure}

Figure \ref{fig:z=1} (a) shows $\q$ dependence of $\lambda_{\q}$ at
$T=30$meV in the FeSe model for $z=1$. We see that the $\q=\bm{0}$ FO order
dominates over $\q=(0,\pi)$ AFB order. This result is confirmed by the
$T$-dependences of $\lambda_{\q}$ for $\bm{q}=(0,\pi)$ and
$\bm{q}=\bm{0}$ shown in Fig. \ref{fig:z=1} (b), which are similar to the
results in main text for $z^{-1}=1.6$ shown in Fig. \ref{fig:nematic} (d).
Thus, AFB order is absent in the FeSe model even for $z=1$.
The qualitative results do not depend on the value of $z$.

%%%%%%%%%%%%%%%%%%%%%%%%%%%%%%%%%
%%%%%%%%%%%%%%%%%%%%%%%%%%%%%%%%%

\color{black}
\subsection{E: Conserving approximation}
In the main text, we employ the RPA, in which the self-energy is not included in the spin (charge)
susceptibility ${\hat \chi}^{s(c)}$ and kernel function $\hat{K}^{\bm{q}}(k,k')$. For this
reason, the DW equation in the RPA violates the
conserving-approximation (CA) formalism of Baym and Kadanoff \cite{Baym,Tremblay}.
The great merit of CA is
that it rigorously satisfies the macroscopic conservation laws. This merit is important to avoid unphysical results. Here, we first calculate the one-loop self-energy
by using the fluctuation exchange (FLEX) approximation
\cite{S-Onari-form,FLEX}. 
Next, we solve the DW equation including the FLEX self-energy to satisfy the CA formalism.

The FLEX self-energy ($C_4$) is given by $\hat{\Sigma}(k)=\frac{T}{N}\sum_q\hat{V}^\Sigma(q)\hat{G}(k-q)$, where
${\hat G}(k)=[(i\e_n-\mu){\hat1}-{\hat{h}}^0(\k)-\hat{\Sigma}(k)]^{-1}$  is the Green
function with the self-energy,
and $\hat{V}^\Sigma$ is the interaction matrix
for the self-energy. 
$\hat{V}^\Sigma$ is given as
\begin{eqnarray}
\hat{V}^\Sigma&=&\frac32 {\hat \Gamma}^s{\hat \chi^s}(q){\hat \Gamma}^s
+\frac12 {\hat \Gamma}^c{\hat \chi^c}(q){\hat \Gamma}^c\nonumber\\
&&-\frac12 \left[{\hat \Gamma}^c{\hat\chi}^0(q){\hat \Gamma}^c
+{\hat \Gamma}^s{\hat\chi}^0(q){\hat \Gamma}^s\right.\nonumber\\
&&-\frac14 \left.({\hat \Gamma}^s+{\hat \Gamma}^c){\hat\chi}^0(q)
({\hat \Gamma}^s+{\hat \Gamma}^c) \right].
\end{eqnarray}
We solve $\hat{\Sigma}$, $\hat{G}$, and $\hat{\chi}^{s(c)}$
self-consistently. The effect of the renormalization factor
$z$
is given by the self-energy $\hat{\Sigma}$ in this framework.
By introducing the obtained functions, we improve the
kernel of the DW equation in Eq. (\ref{eqn:K}) and (\ref{eqn:S-K}) and solve the symmetry-breaking
self-energy (form factor) in the
framework of CA.

%%%%%%%%%%%%%%%%%%%%%%%%%%%%%%%%%
\begin{figure}[!htb]
\includegraphics[width=.99\linewidth]{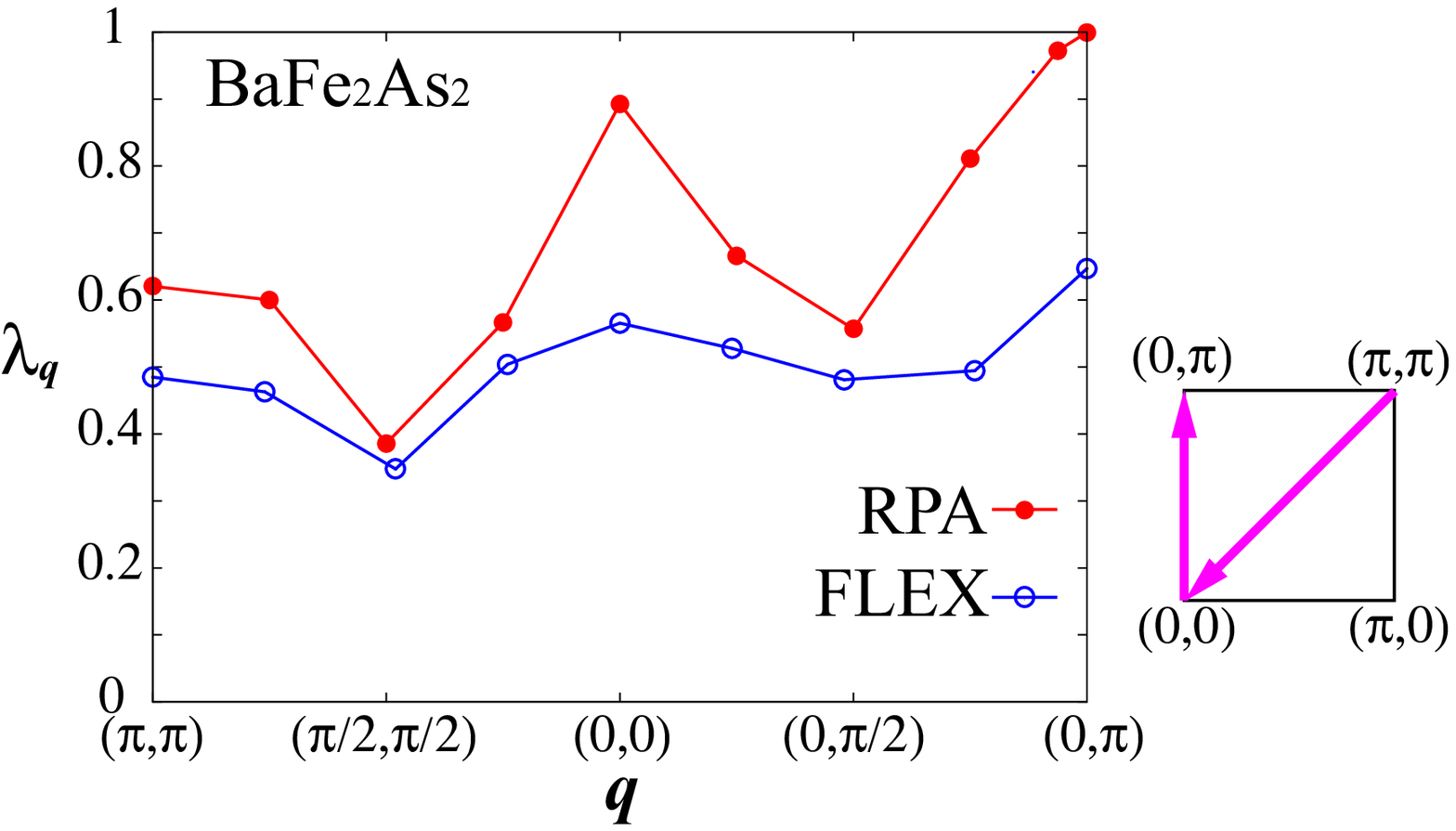}
\caption{
$\bm{q}$ dependences of $\lambda_{\q}$ obtained using the RPA at $T=T^*$ $(\a_s=0.951)$ and the
 FLEX approximation at $T=5$meV $(\a_s=0.945)$ in the BaFe$_2$As$_2$ model. The $\q$ path is shown by arrows in the right panel.
}
\label{fig:FLEX}
\end{figure}
%%%%%%%%%%%%%%%%%%%%%%%%%%%%%%%%%

Figure \ref{fig:FLEX} shows the $\q$ dependences of $\lambda_\q$ obtained using
the RPA at $T=T^*=32.4$meV (spin Stoner factor $\a_s=0.951$) and the FLEX approximation at
$T=5$meV $(\a_s=0.945, r=0.680)$. The parameter $\a_s$ is a
univariate function of the reduction parameter $r$ in Eq. (\ref{eqn:Ham}), and $1-\a_s$ represents the distance from the magnetic
critical point.
In the numerical study, we take $N=100\times100$ $\k$-meshes and
$1024$ Matsubara frequencies. The obtained mass-enhancement
factor $z^{-1}=m^*/m\sim 5$ by the self-energy is consistent with experimental value.
The result of the FLEX approximation is similar to that of RPA employed in
the main text. The AFB order $f^{(0,\pi)}$ is dominant over the
FO order $f^{\bm{0}}$. Thus, the results in the main text are
verified in the
CA framework. 
The obtained value $\lambda_{(0,\pi)}\sim 0.6$ in the CA is still large,
and it will reach unity for larger $r$ or at lower $T$.

\subsection{F: Pairing interaction by the AFB fluctuations}
\color{black}
Here, we discuss the mechanism of superconductivity. In our previous
paper \cite{S-Kontani-Saito-Onari}, we proposed the mechanism by which
the $s_{++}$-wave is mediated by AF orbital (AFO) fluctuations. Therein, we employed
the $\k$-independent form
factor 
$\hat{f}^{\q}(\k)=\hat{O}_\Gamma$ $(\Gamma=xz,yz,xy)$, where
$\hat{O}_\Gamma$ is the local charge quadrupole operator. 
The obtained form factor in the main text is mainly the $\k$-dependent AFB
order, which is different from the AFO order.

\begin{figure}[!htb]
\includegraphics[width=.99\linewidth]{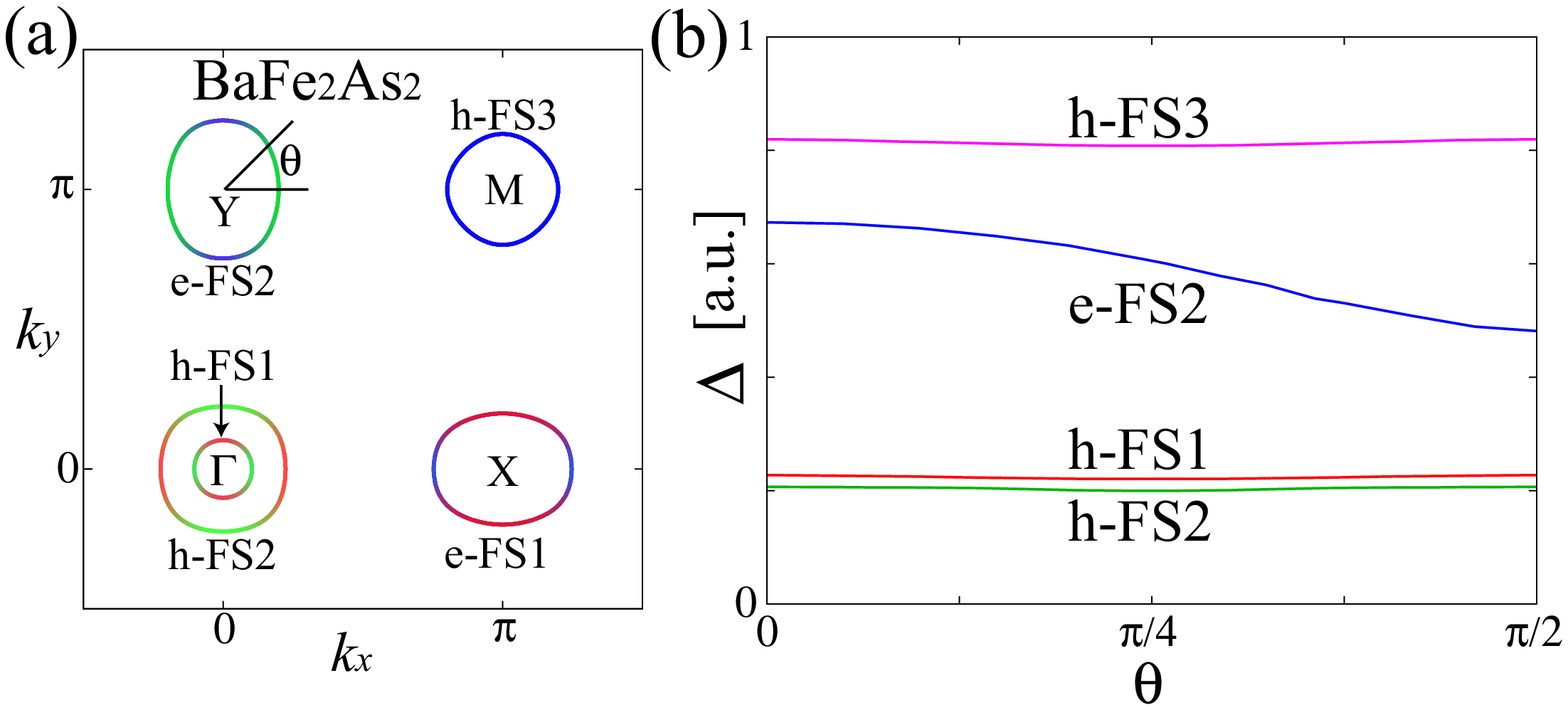}
\caption{
(a) FSs of the BaFe$_2$As$_2$ model in the unfolded zone, where
$\theta$ denotes the azimuthal angle with respect to
 the $x$ axis on each FS.
(b) Superconducting
 gap function at $T=30$meV given by the DW fluctuations as a function of
 $\theta$.
}
\label{fig:super}
\end{figure}

In the
following, we show that the $s_{++}$-wave is realized because of the AFB fluctuations
by using the $\k$-dependent and $\q$-dependent form factor obtained
microscopically from the DW equation.
The pairing interaction $\hat{V}^{\rm DW}$ originating from the DW
susceptibility $\chi_{\rm nem}(\q)\propto
(1-\lambda_{\q})^{-1}$ \cite{S-Onari-B2g} is expanded by the form factor as
\begin{eqnarray}
\!\!\!\!\!\!\!\!\!\!\! V^{\rm
 DW}_{l,l';m,m'}(k,k')&=&\nonumber\\
&&\!\!\!\!\!\!\!\!\!\!\!\!\!\!\!\!\!\!\!\!\!\!\!\!\!\!\!\!\!\!\!\!\!\!\!\!\!\!\!\sum_{\bm{q}=\bm{0},\bm{Q},\bm{Q}'}\!\!\!\!\!
 f^{\bm{q}}_{l,l'}(\bm{k'})\frac{\alpha^{\q}}{1+\xi_{\q}^2(\bm{k}-\bm{k'}-\bm{q})^2}
f^{\bm{q}*}_{m,m'}(-\bm{k'}-\q), \label{DW-SC}
\end{eqnarray}
where $\Q=(0,\pi)$, $\Q'=(\pi,0)$, $\alpha^{\q}=\bar{I}^{\q}/(1-\lambda_{\q})$, and $\xi_{\q}$ denotes
the correlation length for the DW state with wave vector $\q$. We put $\xi_{\q}=1.0$.
$\bar{I}^{\q}$ is the mean value of
$\hat{I}^{\q}$ in the basis of
$\hat{f}^{\q}(\bm{k})$ at the lowest Matsubara frequency. It is given as
\begin{equation}
\bar{I}^{\bm{q}}=\frac{\sum_{\k,\bm{k'}}\sum_{l,l',m,m'}f^{\bm{q}*}_{l,l'}(\bm{k})I^{\bm{q}}_{l,l';m,m'}(\bm{k},\bm{k'})f^{\q}_{m,m'}(\bm{k'})}{\left[\sum_{\bm{k}}\sum_{l,l'}|f^{\bm{q}}_{l',l}(\k)|^2\right]^2}.
\end{equation}
We obtain $\bar{I}^{\bm{0}}\sim 9$eV in the BaFe$_2$As$_2$ model.
Because of the large $\bar{I}^{\bm{q}}/U\sim 6$, the AFB
fluctuations yield a large attractive pairing
interaction. The obtained enhancement of $\bar{I}^{\bm{q}}$ is
consistent with the large charge-channel U-VC at low energies in the 
beyond-Migdal-Eliashberg theory in Refs. \cite{S-Onari-SCVCS,S-Tazai}.
We solve the following Eliashberg equation, which includes the RPA pairing
interaction and the DW fluctuation pairing interaction $\hat{V}^{\rm
DW}$  with the cutoff energy $W_c=0.02$eV:
{\small
\begin{eqnarray}
\lambda_{\rm
 SC}\Delta_{l,m}(k)&=&\frac{T}{N}\sum_{k'}\sum_{l',l'',m',m''}\left[-\frac{3}{2}\hat{V}^s(k-k')+\frac{1}{2}\hat{V}^c(k-k')\right.\nonumber\\
&&\left.+\hat{V}^{\rm DW}(k,k')+\frac{1}{2}(\hat{\Gamma}^c-\hat{\Gamma}^s)\right]_{l,l';m',m}
 \nonumber \\
&&\times G_{l',l''}(k')G_{m',m''}(-k')\Delta_{l'',m''}(k').
\end{eqnarray}
}
In order to concentrate on the FSs, we define $\theta$ as the azimuthal
angle $\theta$ with respect to the $x$ axis on each FS shown in
Fig. \ref{fig:super}(a).  
Figure \ref{fig:super}(b) shows the
 derived $s_{++}$-wave 
gap function $\Delta$ as a function of the $\theta$ on each FS.
We find that the
fullgap $s_{++}$ wave is obtained because the obtained
AFB fluctuations with large inter and intra (2--4) orbital components
yield a sizable inter-FS
attractive pairing interaction. Thus, we find that the AFB fluctuations
with a non-local form factor is a novel mechanism
of the $s_{++}$ wave. The present AFB-fluctuation mechanism is a natural extension of
the beyond-Migdal-Eliashberg theory with U-VC developed in Refs. \cite{S-Onari-SCVCS,S-Tazai}.
In future publications, we will discuss the details of results and the mechanism of
superconductivity based on the DW fluctuations.
\color{black}
%%%%%%%%%%%%%%%%%%%%%%%%%%%%%%%%%
%%%%%%%%%%%%%%%%%%%%%%%%%%%%%%%%%

%%%%%%%%%%%%%%%%%%%%%%%%
%references
%%%%%%%%%%%%%%%%%%%%%%%%

\end{document}